\documentstyle[12pt]{article}
\newcommand{\beq}{\begin{equation}}
\newcommand{\eeq}{\end{equation}}
\newcommand{\ba}{\begin{eqnarray}}
\newcommand{\ea}{\end{eqnarray}}

\newcommand{\dsl}
  {\kern.06em\hbox{\raise.15ex\hbox{$/$}\kern-.56em\hbox{$\partial$}}}

\newcommand{\eeqarr}{\end{eqnarray}}
\newcommand{\ZZ}{{\rm \kern 0.275em Z \kern -0.92em Z}\;}
\begin{document}
\begin{titlepage}
\begin{center}
\vspace*{0.5cm}

{\Huge Four Point Functions in} \\
\vspace*{0.7cm}
{\Huge the $SL(2,R)$ WZW Model}
\\
\vspace*{1.5cm}
{\large Pablo 
Minces$^{a,}$\footnote{minces@iafe.uba.ar} and
Carmen N\'u\~nez$^{a,b,}$\footnote{carmen@iafe.uba.ar}}\\
\vspace*{1.5cm}
$^{a}$Instituto de Astronom\'{\i}a y F\'{\i}sica del Espacio 
(IAFE),\\
C.C.67 - Suc. 28, 1428 Buenos Aires, Argentina.\\
\vspace*{0.4cm}
$^{b}$Physics Department, University of Buenos Aires,\\
Ciudad Universitaria, Pab. I, 1428 Buenos Aires, Argentina.\\
\vspace*{1.5cm}

\end{center}
\begin{abstract}
We consider winding conserving four point functions in the  $SL(2,R)$ 
WZW model 
for states in arbitrary spectral flow sectors. We compute the leading 
order contribution to the expansion of the amplitudes in powers of the 
 cross ratio of the four points on the  worldsheet, 
both in the $m-$ and  $x-$basis,
with at least one state in the spectral flow image of the highest weight 
discrete representation. We also perform certain consistency check on the 
winding conserving three point functions.
\end{abstract}

\end{titlepage}

The $SL(2,R)$ WZW model describes strings moving in AdS$_3$ and
has important applications 
to gravity and black hole physics in two and three dimensions.
It is also an
interesting subject by itself, as a step beyond the well known rational 
conformal field theories. Actually,
unlike string propagation in compact
target spaces, where 
the spectrum is in general 
discrete and
the model can then be studied using algebraic methods,
 the analysis of the worldsheet theory in the non-compact 
AdS$_3$ background
requires the use of more intricate analytic techniques.

String theory on AdS$_3$ contains different sectors characterized by an integer
number $w$, the spectral flow parameter or winding number \cite{malda1}. 
The {\it short} string
sectors correspond to maps from the worldsheet to a compact 
region in AdS$_3$ and the states in this sector belong to
discrete representations of $SL(2,R)$ with spin $j\in$ {\bf R} and
 unitarity bound $\frac 12 < j < \frac{k-1}2$. Other sectors contain 
{\it long}
strings at infinity, near the boundary of spacetime, described by continuous 
representations of $SL(2,R)$ with spin $j=\frac 12 + is$, $s\in$ {\bf R}.
Several correlation functions have been computed in various sectors
\cite{malda3, hmn}. In particular, four point functions of $w=0$ states
were computed in \cite{malda3} analytically continuing previous results in
the $SL(2,C)/SU(2)$ coset model \cite{tesch2}
which corresponds to the Euclidean $H_3$ background. In this letter we consider
four point functions of states in arbitrary $w$ sectors, a crucial ingredient
 to determine the consistency of the theory through factorization.

Different basis have been used in the literature to compute correlation
functions in this theory. Vertex operators and expectation values for the 
spectral flow representations were constructed in \cite{malda1, malda3}
in the $m-$basis, where the generators $(J_0^3,\bar J_0^3 )$ of the global
$SL(2,C)$ isometry are diagonalized. The $m-$basis has the advantage that
all values of $w$ can be treated simultaneously. In particular, 
 all winding
conserving $N-$point functions have the same coefficient in this basis, 
for a given $N$, 
and they differ only in the worldsheet coordinate
dependence \cite{malda3, ribault} which reflects the change in the conformal
weight of the states
\beq
\Delta(j) \rightarrow \Delta^w (j) = \Delta(j) - wm - \frac k4 w^2 \quad ,
\label{flow}
\eeq
where $\Delta(j)=-\frac{j(j-1)}{k-2}$ is the dimension of the unflowed 
operators.

Alternatively, the $x-$basis refers to the $SL(2,R)$ isospin parameter
which can be interpreted as the coordinate of the boundary in the context 
of the AdS/CFT correspondence. The operators $\Phi_{j}(x,\bar x)$ in the 
$x-$basis
and $\Phi_{j;m,\bar m}(z, \bar z)$ 
in the $m-$basis are related by the following 
transformation
\beq
\Phi_{j; m, \bar m}(z,{\bar z}) = \int \frac{d^2 x}{|x|^2}x^{j-m}\bar x^{j
- \bar m} \Phi_{j}(x,\bar x;z,{\bar z}) \; .
\label{max}
\eeq
Finally,  the $\mu -$basis was found convenient to relate correlation functions
in Liouville and $SL(2,C)/SU(2)$ models \cite{ribault, rt}.

In this letter we
extend the results for the four point function of unflowed states 
in $SL(2,R)$ given in \cite{malda3} to the case of winding conserving 
four point functions for states in arbitrary spectral flow sectors.
This is accomplished by transforming the $x-$basis expression found in
\cite{tesch2} to the $m-$basis in order to
exploit the fact that the coefficient of all winding conserving correlators 
is the same (for a given number of external states).\footnote{This fact 
is shown in Appendix B for the winding conserving three point function
with states in the sectors $w=+1, -1$, and $0$.} 
In order to 
simplify the calculations we consider first the four point function in which
 one of the original unflowed operators is a highest weight and then 
analyze the more general case in which it is replaced by a global 
$SL(2,R)$ descendant. Finally we transform the result back to the 
$x-$basis. 

Actually the explicit expression computed in \cite{tesch2} 
and further analyzed in \cite{malda3} corresponds to the leading order 
in the expansion of the four point function in powers of the 
cross ratio of the worldsheet coordinates. It was pointed out in 
\cite{tesch2} that higher orders in the expansion may be determined 
iteratively once the lowest order is fixed as the initial condition. We 
will further discuss this topic for four point functions involving
spectral flowed states. 

Let us start recalling the result for the four point function of unflowed 
states originally computed for the $SL(2,C)/SU(2)$ model in \cite{tesch2} 
and later analytically continued to $SL(2,R)$  in 
\cite{malda3}, namely
\ba
&&\left<\Phi_{j_{1}}(x_{1},z_{1})
\Phi_{j_{2}}(x_{2},z_{2})\Phi_{j_{3}}(x_{3},z_{3})
\Phi_{j_{4}}(x_{4},z_{4})\right>\nonumber\\
&& \qquad =\;\int dj\; C(j_{1},j_{2},j)\; B(j)^{-1}
C\left(j,j_{3},j_{4}\right) {\cal F}(z,x)\; {\bar{\cal F}}({\bar z},{\bar
x})
\nonumber\\ && \qquad\times\;
|x_{43}|^{2(j_{1}+j_{2}-j_{3}-j_{4})}
|x_{42}|^{-4j_{2}}|x_{41}|^{2(j_{2}+j_{3}-j_{1}-j_{4})}
|x_{31}|^{2(j_{4}-j_{1}-j_{2}-j_{3})}
\nonumber\\
&& \qquad\times\;
|z_{43}|^{2(\Delta_{1}+\Delta_{2}-\Delta_{3}-\Delta_{4})}
|z_{42}|^{-4\Delta_{2}}|z_{41}|^{2(\Delta_{2}+\Delta_{3}-\Delta_{1}-
\Delta_{4})}
|z_{31}|^{2(\Delta_{4}-\Delta_{1}-\Delta_{2}-\Delta_{3})}
\; ,
\label{qa}
\ea
where the integral is over $j=\frac{1}{2}+is$ with $s$ a positive real 
number. Here $B$ and $C$ are the coefficients corresponding to the 
two and three point functions of unflowed operators respectively 
(see \cite{malda3} for the explicit expression in our conventions), and 
${\cal F}$ is a function of the cross ratios 
$ z=\frac{z_{21}z_{43}}{z_{31}z_{42}} , ~
x=\frac{x_{21}x_{43}}{x_{31}x_{42}}$,
with a similar expression for the antiholomorphic part. The function 
${\cal F}$ is obtained by requiring (\ref{qa}) to be a solution of 
the Knizhnik-Zamolodchikov (KZ) equation \cite{zamo2,zamo3}. 

Expanding ${\cal F}$ in powers of $z$ as follows
\ba
{\cal
F}(z,x)=z^{\Delta_{j}-\Delta_{1}-\Delta_{2}}\; x^{j-j_{1}-j_{2}}
\sum_{n=0}^{\infty}f_{n}(x)z^{n}\; ,
\label{expan}
\ea
the lowest order $f_{0}$ is determined to be a solution of the 
standard hypergeometric equation thus giving two linearly independent 
solutions
\ba
&&{_{2}F_{1}}(j-j_{1}+j_{2},j+j_{3}-j_{4},2j;x)\; ,\nonumber\\
&{\rm or}& x^{1-2j}{_{2}F_{1}}(1-j-j_{1}+j_{2},1-j+j_{3}-j_{4},2-2j;x)\; .
\nonumber
\ea
Taking into account both the holomorphic and antiholomorphic parts, 
the unique monodromy invariant combination is of the form \cite{malda3}
\ba
&&|{\cal F}(z,x)\;|^2\;=\;
|z|^{2(\Delta_{j}-\Delta_{1}-\Delta_{2})}\; 
|x|^{2(j-j_{1}-j_{2})}
{\Bigg 
\{}\;{\bigg 
|}\;{_{2}F_{1}}(j-j_{1}+j_{2},j+j_{3}-j_{4},2j;x)\;{\bigg 
|}^{2}\nonumber\\ 
&& \qquad
\quad\qquad
+\; \lambda \;
{\bigg 
|}\;x^{1-2j}{_{2}F_{1}}(1-j-j_{1}+j_{2},1-j+j_{3}-j_{4},2-2j;x)\;{\bigg 
|}^{2}\; 
{\Bigg \}}
+\;\cdots \; ,\nonumber\\
\label{qaa}
\ea
where the ellipses denote higher orders in $z$ and
\ba
\lambda=\;-\;\frac{\gamma(2j)^{2}\gamma(-j_{1}+j_{2}-j+1)
\gamma(j_{3}-j_{4}-j+1)}{(1-2j)^{2}\gamma(-j_{1}+j_{2}+j)
\gamma(j_{3}-j_{4}+j)}\; ,
\nonumber
\ea 
with $\gamma(a)\equiv\frac{\Gamma(a)}{\Gamma(1-a)}$. Higher orders in 
(\ref{expan}) are determined 
iteratively by the KZ equation starting from $f_{0}$ 
as the initial condition \cite{tesch2}.

Now we perform the transformation of (\ref{qa}) to the $m-$basis 
with the 
solution (\ref{qaa}). 
Let us consider first the case in which one of the operators
in the four point function is a
 highest weight,\footnote{The more general 
case where the highest weight 
is replaced by a global $SL(2,R)$ descendant is analyzed below.} say 
$\Phi_{j_{1}}$,  
and look at the contribution of the first 
term in the r.h.s. of (\ref{qaa}) (the second term 
 will be considered later). The $x_i$ dependence is given by
\ba
|K(x_{i},j_{i},j)|^2={\bigg |} 
x_{43}^{j_{1}+j_{2}-j_{3}-j_{4}}x_{42}^{-2j_{2}}
x_{41}^{j_{2}+j_{3}-j_{1}-j_{4}}x_{31}^{j_{4}-j_{1}-j_{2}-j_{3}}
x^{j-j_{1}-j_{2}}{_{2}F_{1}}(a,b,c;x)\; {\bigg |}^2\; ,
\label{01}
\ea
where
$a=j-j_{1}+j_{2}$, $b=j+j_{3}-j_{4}$ and $ c=2j$.

We have to evaluate 
the residue of the pole at $j_{1}=-m_{1}=-\bar m_1$ 
in the $x_1$ integral transform of (\ref{01}). This is obtained
 applying
the following operator 
\ba
\int d^{2}x_{2}\; d^{2}x_{3}\; d^{2}x_{4}\; x_{2}^{j_{2}-m_{2}-1}
{\bar x}_{2}^{j_{2}-{\bar m}_{2}-1}
x_{3}^{j_{3}-m_{3}-1}
{\bar x}_{3}^{j_{3}-{\bar m}_{3}-1}x_{4}^{j_{4}-m_{4}-1}
{\bar x}_{4}^{j_{4}-{\bar m}_{4}-1}
 \lim_{x_{1},{\bar x_{1}}\rightarrow 
\infty}\; x_{1}^{2j_{1}}{\bar x}_{1}^{2 j_{1}} \; ,
\nonumber
\ea
where we are using (\ref{max}).

Noticing that
\ba
&& \lim_{x_{1}\rightarrow \infty}\; [x_{1}^{2j_{1}}
K(x_{i},j_{i},j)] 
 =\;\nonumber\\
&&\qquad\qquad\qquad x_{4}^{j_{1}-j_{2}-j_{3}-j_{4}}
\left(1-\frac{x_{3}}{x_{4}}\right)^{j-j_{3}-j_{4}}
\left(1-\frac{x_{2}}{x_{4}}\right)^{j_{1}-j_{2}-j}
{_{2}F_{1}}\left(a,b,c;\frac{1-\frac{x_{3}}{x_{4}}}
{1-\frac{x_{2}}{x_{4}}}\right)\; ,\nonumber
\ea
and
performing the change of variables 
$x_{2}\rightarrow
\frac{x_{2}}{x_{4}}$, $x_{3}\rightarrow
\frac{x_{3}}{x_{4}}$, 
we may write
\ba
&& \lim_{x_{1},{\bar x}_{1}\rightarrow \infty}
\int d^{2}x_{2}\; d^{2}x_{3}\; d^{2}x_{4}\; x_{2}^{j_{2}-m_{2}-1}
{\bar x}_{2}^{j_{2}-{\bar m}_{2}-1}
x_{3}^{j_{3}-m_{3}-1}
{\bar x}_{3}^{j_{3}-{\bar m}_{3}-1}x_{4}^{j_{4}-m_{4}-1}
{\bar x}_{4}^{j_{4}-{\bar m}_{4}-1}\nonumber\\ &&\qquad\qquad
\times \;
 |x_{1}^{2j_{1}} 
K(x_{i},j_{i},j)|^2\sim
V_{conf}\;\delta^{2}(m_{1}+m_{2}+m_{3}+m_{4})
\;\Omega (j,j_{i},m_{i},{\bar m}_{i})\; ,\nonumber\\
\label{omega}
\ea
where
\ba
&&\Omega (j,j_{i},m_{i},{\bar m}_{i})\nonumber\\
&&\equiv \int d^{2}x_{2}\; d^{2}x_{3}\;
x_{2}^{j_{2}-m_{2}-1}{\bar x}_{2}^{j_{2}-{\bar m}_{2}-1}
x_{3}^{j_{3}-m_{3}-1}{\bar x}_{3}^{j_{3}-{\bar m}_{3}-1}
|1-x_{2}|^{2(j_{1}-j_{2}-j)}|1-x_{3}|^{2(j-j_{3}-j_{4})}
\nonumber\\ && \qquad\qquad\qquad\qquad\qquad\qquad\qquad
\times\;
{_{2}F_{1}}\left(a,b,c;\frac{1-x_{3}}{1-x_{2}}\right)
{_{2}F_{1}}\left(a,b,c;\frac{1-{\bar x}_{3}}{1-{\bar x}_{2}}\right)\; ,
\label{int}
\ea
and the $\delta$-function comes from the  $x_{4}$ integral. The 
factor $V_{conf}$ 
is the volume of 
the conformal group of $S^{2}$, and it arises 
from the fact that we are looking at
the residue of the
pole at $j_{1}=-m_{1}=-\bar m_1$ (see \cite{malda3}).

In order to compute $\Omega$, we begin by considering the 
following integral
\ba
I\equiv \int d^{2}y\; y^{p}{\bar y}^{q} 
(1-y)^{r}(1-{\bar y})^{s}{_{2}F_{1}}(a,b,c;t(1-y))
{_{2}F_{1}}(a,b,c;{\bar t}(1-{\bar y}))\; .
\nonumber
\ea
We approach it by first redefining  
variables and 
integration contours.\footnote{This is similar to the 
computation of the integral $\int d^{2}y\; 
|y^{\alpha}(1-y)^{\beta} |^{2}$ in \cite{dots}.} 
Let us define $y_{1}$ and $y_{2}$ 
through
$y=y_{1}+iy_{2},~ {\bar y}=y_{1}-iy_{2}$,
and then perform the scaling $y_{2}\rightarrow -ie^{2i\epsilon}y_{2}$, 
where $\epsilon$ is a small positive number. Thus we may write
\ba
&& I\sim \int_{-\infty}^{\infty}dy_{1}
\int_{-\infty}^{\infty}dy_{2}\; 
(y_{1}-y_{2}+2i\epsilon y_{2})^{p}(y_{1}+y_{2}-2i\epsilon y_{2})^{q}
(1-y_{1}+y_{2}-2i\epsilon 
y_{2})^{r}\nonumber\\ &&\qquad\qquad\qquad\qquad\quad \times\;
(1-y_{1}-y_{2}+2i\epsilon 
y_{2})^{s}{_{2}F_{1}}(a,b,c;t(1-y_{1}+y_{2}-2i\epsilon y_{2}))
\nonumber\\ &&\qquad\qquad\qquad\qquad\qquad\qquad\qquad\qquad\qquad 
\times\;
{_{2}F_{1}}(a,b,c;{\bar t}(1-y_{1}-y_{2}+2i\epsilon 
y_{2}))\; .
\nonumber
\ea
It is convenient to introduce 
$y_{\pm}\equiv y_{1}\pm y_{2}$
so that the integral can be rewritten as
\ba
&& I\sim \int_{-\infty}^{\infty}dy_{+}\int_{-\infty}^{\infty}dy_{-}
\;
[y_{+}-i\epsilon(y_{+}- y_{-})]^{q}
[1-y_{+}+i\epsilon(y_{+}- y_{-})]^{s}
\nonumber\\ &&\qquad\qquad\qquad\qquad\qquad\qquad \times\;
{_{2}F_{1}}(a,b,c;{\bar t}[1-y_{+}+i\epsilon(y_{+}- 
y_{-})])
\nonumber\\ &&\quad\qquad\qquad\qquad\qquad\times \;
[y_{-}+i\epsilon(y_{+}- y_{-})]^{p}
[1-y_{-}-i\epsilon(y_{+}- y_{-})]^{r}
\nonumber\\ &&\qquad\qquad\qquad\qquad\qquad\qquad\times\;
{_{2}F_{1}}(a,b,c;t[1-y_{-}-i\epsilon(y_{+}-y_{-})])\; .
\label{rt}
\ea
Here the $\epsilon$ terms define the way we go around the singular points 
(see \cite{dots} for details). We decompose the integration interval 
for 
$y_{+}$ into $(-\infty,0)\cup (0,1)\cup (1,\infty)$. 
The only non-vanishing contribution comes from the 
interval $y_{+}\in(0,1)$, 
since the integration 
contours for $y_{-}$ can be deformed to infinity
in the other two intervals, and we assume that the 
integrals are convergent. For
 the non-vanishing contribution, we 
 deform the integration contour to 
$y_{-}$ in $(1,\infty)$.

In this way the $m$ and $\bar m$
contributions are 
factorized as 
follows
\ba
&& I= sin(\pi r)\int_{0}^{1}dy_{+}\;
y_{+}^{q}
(1-y_{+})^{s}
{_{2}F_{1}}(a,b,c;{\bar t}(1-y_{+}))
\nonumber\\ &&\qquad\qquad\times \;
\int_{1}^{\infty}dy_{-}\;
y_{-}^{p}
(1-y_{-})^{r}
{_{2}F_{1}}(a,b,c;t(1-y_{-}))\; .
\nonumber
\ea
Now changing  variable $y_{-}\rightarrow \frac{1}{y_{-}}$ we 
arrive at
\ba
&& I= (-1)^{r}sin(\pi r)\int_{0}^{1}dy_{+}\;
y_{+}^{q}
(1-y_{+})^{s}
{_{2}F_{1}}(a,b,c;{\bar t}(1-y_{+}))
\nonumber\\ &&\qquad\qquad\qquad\times \;
\int_{0}^{1}dy_{-}\;
y_{-}^{-p-r-2}
(1-y_{-})^{r}
{_{2}F_{1}}\left(a,b,c;t\frac{y_{-}-1}{y_{-}}\right)\; .
\label{sd}
\ea

Consider now the integral in (\ref{int}). Using twice the result above we 
find
\ba
&&\Omega(j,j_{i},m_{i},{\bar m}_{i})=\;
\Omega_{-}(j,j_{i},m_{i}) \; \Omega_{+}(j,j_{i},{\bar m}_{i}) \; ,
\label{ns}
\ea
where
\ba
&&\Omega_{-}(j,j_{i},m_{i}) = (-1)^{j_{1}-j_{2}-j_{3}-j_{4}}sin[\pi 
(j_{1}-j_{2}-j)]
sin[\pi (j-j_{3}-j_{4})]\nonumber\\
&&\qquad\qquad\times
\int_{0}^{1}du_{-}\int_{0}^{1}dv_{-}\; u_{-}^{j_{4}-j+m_{3}-1}
(1-u_{-})^{j-j_{3}-j_{4}}v_{-}^{j-j_{1}+m_{2}-1}(1-v_{-})^{j_{1}-j_{2}-j}
\nonumber\\
&&
\qquad\qquad\qquad\qquad\qquad\qquad\qquad
\times\;
{_{2}F_{1}}\left(a,b,c;\frac{v_{-}}{u_{-}}\;\frac{1-u_{-}}{1-v_{-}}\right)\; ,
\nonumber
\ea
 
\ba
&&\Omega_{+}(j,j_{i},{\bar m}_{i}) =
\int_{0}^{1}du_{+}\int_{0}^{1}dv_{+}\; u_{+}^{j-j_{3}-j_{4}}
(1-u_{+})^{
j_{3}-{\bar
m}_{3}-1}v_{+}^{j_{1}-j_{2}-j}
(1-v_{+})^{
j_{2}-{\bar m}_{2}-1}
\nonumber\\ &&\qquad\qquad\qquad\qquad\qquad\qquad\qquad\qquad\times\;
{_{2}F_{1}}\left(a,b,c;\frac{u_{+}}{v_{+}}\right)\; , 
\label{k1}
\ea
and in the last integral we have performed the additional 
change of variables $u_{+}\rightarrow 1-u_{+}$, $v_{+}\rightarrow 1-v_{+}$.

Now notice that the roles of $y_+$ and $y_-$ can be exchanged in the 
manipulations leading from (\ref{rt}) to (\ref{sd}). In this case one 
would arrive at an equivalent expression 
\ba
&& I= (-1)^{s}sin(\pi s)\int_{0}^{1}dy_{+}\;
y_{+}^{-q-s-2}
(1-y_{+})^{s}
{_{2}F_{1}}\left(a,b,c;{\bar 
t}\frac{y_{+}-1}{y_{+}}\right)
\nonumber\\ &&\qquad\qquad\qquad\qquad\times \;
\int_{0}^{1}dy_{-}\;
y_{-}^{p}
(1-y_{-})^{r}
{_{2}F_{1}}\left(a,b,c;t(1-y_{-})\right)\; ,
\nonumber
\ea
which leads to  
\ba
&&\Omega_{-}(j,j_{i},m_{i}) =
\int_{0}^{1}du_{-}\int_{0}^{1}dv_{-}\; u_{-}^{j-j_{3}-j_{4}}
(1-u_{-})^{
j_{3}-m_{3}-1}v_{-}^{j_{1}-j_{2}-j}
(1-v_{-})^{j_{2}-m_{2}-1}
\nonumber\\ &&\qquad\qquad\qquad\qquad\qquad\qquad\qquad\qquad\times\;
{_{2}F_{1}}\left(a,b,c;\frac{u_{-}}{v_{-}}\right)\; .
\label{k2}
\ea

Comparing Eqs.(\ref{k1}) and (\ref{k2}) we are now able to write 
$\Omega$ in a form which is {\it manifestly} symmetric with respect to 
$m_{i}$ and ${\bar m}_{i}$, namely
\ba
&&\Omega =
\int_{0}^{1}du_{-}\int_{0}^{1}dv_{-}\; u_{-}^{j-j_{3}-j_{4}}
(1-u_{-})^{
j_{3}-m_{3}-1}v_{-}^{j_{1}-j_{2}-j}
(1-v_{-})^{
j_{2}- m_{2}-1}
\nonumber\\ &&\qquad\qquad\qquad\qquad\qquad\qquad\qquad\qquad\qquad
\qquad\quad\times\;
{_{2}F_{1}}\left(a,b,c;\frac{u_{-}}{v_{-}}\right)\nonumber\\
&&\qquad\qquad\times \;
\int_{0}^{1}du_{+}\int_{0}^{1}dv_{+}\; u_{+}^{j-j_{3}-j_{4}}
(1-u_{+})^{
j_{3}-{\bar m}_{3}-1}v_{+}^{j_{1}-j_{2}-j}
(1-v_{+})^{
j_{2}-{\bar m}_{2}-1}
\nonumber\\ &&\qquad\qquad\qquad\qquad\qquad\qquad\qquad\qquad\qquad
\qquad\quad\times\;
{_{2}F_{1}}\left(a,b,c;\frac{u_{+}}{v_{+}}\right)\; .
\label{tq}
\ea

Therefore, the problem of computing $\Omega$ reduces to that of solving 
the following integral
\ba
\Sigma =\int_{0}^{1}du\int_{0}^{1}dv\; u^{\alpha}
(1-u)^{\beta}v^{\mu}
(1-v)^{\nu}
{_{2}F_{1}}\left(a,b,c;\frac{u}{v}\right)\; .
\nonumber
\ea
Now using 
(\ref{m1}) (see the Appendix A) we may write
\ba
\Sigma =\frac{\Gamma(\alpha +1)\Gamma(\beta 
+1)}{\Gamma(\alpha +\beta + 2)}\int_{0}^{1}dv\; v^{\mu}
(1-v)^{\nu}
{_{3}F_{2}}\left(a,b,\alpha +1;c,\alpha +\beta +2;\frac{1}{v}\right)\; ,
\nonumber
\ea
and with the help of (\ref{m2}) and (\ref{m3}) (see the Appendix A) we 
arrive at
\ba
&&\Sigma =\; \Gamma(\alpha +1)\Gamma(\beta
+1)\nonumber\\ && \qquad\quad\times\;
{\Bigg (}\;
\Lambda\left [ \begin{array}{c}
a,a-c+1,a-\alpha -\beta -1,a+\mu +1\\
a-b+1,a-\alpha,a+\mu +\nu +2
\end{array}
\right ]\nonumber\\ && \qquad\qquad +\;
\Lambda\left [ \begin{array}{c}
b,b-c+1,b-\alpha -\beta -1,b+\mu +1\\
b-a+1,b-\alpha,b+\mu +\nu +2
\end{array}
\right ]\nonumber\\ && \qquad\qquad +\;
\Lambda\left [ \begin{array}{c}
\alpha +1,\alpha -c+2,-\beta,\alpha +\mu +2\\
\alpha -a+2,\alpha-b+2,\alpha +\mu +\nu +3
\end{array}
\right ]\;
{\Bigg )}\; ,
\label{lambda}
\ea
where we have defined
\ba
&&\Lambda\left [ \begin{array}{c}
\rho_{1},\rho_{2},\rho_{3},\rho_{4}\\
\sigma_{1},\sigma_{2},\sigma_{3}
\end{array}\right ]\nonumber\\ && \qquad
\equiv\; (-1)^{\rho_{1}}
\frac{\Gamma(1-\sigma_{1})\Gamma(1-\sigma_{2})
\Gamma(1+\rho_{1}-\rho_{2})\Gamma(\rho_{4})\Gamma(\sigma_{3} -\rho_{4})}
{\Gamma(1-\rho_{2})\Gamma(1-\rho_{3})\Gamma(1+\rho_{1}-\sigma_{1})
\Gamma(1+\rho_{1}-\sigma_{2})\Gamma(\sigma_{3})}
\nonumber\\
&&\qquad\qquad\qquad\qquad\qquad\qquad\qquad\qquad\qquad\quad
\times\; {_4 F_3(\rho_{1},\rho_{2},\rho_{3},\rho_{4};
\sigma_{1},\sigma_{2},\sigma_{3};1)} \; .
\nonumber
\ea

Finally, replacing (\ref{lambda}) in (\ref{tq}) we find the 
explicit form of $\Omega$ as follows
\ba
&&\Omega (j,j_{i},m_{i},{\bar m}_{i}) =\; \Gamma(-j_{3}-j_{4}+j+1)^{2}\;
\Gamma(j_{3}-m_{3})\Gamma(j_{3}-{\bar m}_{3})\nonumber\\ 
&& \times \;
{\Bigg (}\;
\Lambda\left [ \begin{array}{c}
-j_{1}+j_{2}+j,\; -j_{1}+j_{2}-j+1,\; -j_{1}+j_{2}+j_{4}+m_{3},\; 1\\
-j_{1}+j_{2}-j_{3}+j_{4}+1,\; -j_{1}+j_{2}+j_{3}+j_{4},\; j_{2}-m_{2}+1
\end{array}
\right ]\nonumber\\ && +\;
\Lambda\left [ \begin{array}{c}
j_{3}-j_{4}+j,\; j_{3}-j_{4}-j+1,\; j_{3}+m_{3},\; 
j_{1}-j_{2}+j_{3}-j_{4}+1\\
j_{1}-j_{2}+j_{3}-j_{4}+1,\; 2j_{3},\; 
j_{1}+j_{3}-j_{4}-m_{2}+1
\end{array}
\right ]\nonumber\\ && +\;
\Lambda\left [ \begin{array}{c}
-j_{3}-j_{4}+j+1,\; -j_{3}-j_{4}-j+2,\; -j_{3}+m_{3}+1,\; 
j_{1}-j_{2}-j_{3}-j_{4}+2\\
j_{1}-j_{2}-j_{3}-j_{4}+2,\; -2j_{3}+2,\; j_{1}-j_{3}-j_{4}-m_{2}+2
\end{array}
\right ]\;
{\Bigg )}
\nonumber\\
&& \times \;
{\Bigg (}\;
\Lambda\left [ \begin{array}{c}
-j_{1}+j_{2}+j,\; -j_{1}+j_{2}-j+1,\; -j_{1}+j_{2}+j_{4}+{\bar m}_{3},\; 
1\\
-j_{1}+j_{2}-j_{3}+j_{4}+1,\; -j_{1}+j_{2}+j_{3}+j_{4},\; 
j_{2}-{\bar m}_{2}+1
\end{array}
\right ]\nonumber\\ && +\;
\Lambda\left [ \begin{array}{c}
j_{3}-j_{4}+j,\; j_{3}-j_{4}-j+1,\; j_{3}+{\bar m}_{3},\;
j_{1}-j_{2}+j_{3}-j_{4}+1\\
j_{1}-j_{2}+j_{3}-j_{4}+1,\; 2j_{3},\;
j_{1}+j_{3}-j_{4}-{\bar m}_{2}+1
\end{array}
\right ]\nonumber\\ && +\;
\Lambda\left [ \begin{array}{c}
-j_{3}-j_{4}+j+1,\; -j_{3}-j_{4}-j+2,\; -j_{3}+{\bar m}_{3}+1,\;
j_{1}-j_{2}-j_{3}-j_{4}+2\\
j_{1}-j_{2}-j_{3}-j_{4}+2,\; -2j_{3}+2,\; j_{1}-j_{3}-j_{4}-{\bar m}_{2}+2
\end{array}
\right ]\;
{\Bigg )} .
\nonumber\\
\label{oomega}
\ea

We still have
to take into account the second term 
in 
the r.h.s. of (\ref{qaa}). However, as pointed out in \cite{malda3}, this 
can be obtained from the first term  by just
replacing $j\rightarrow 1-j$. Therefore the explicit form of the four 
point function of 
unflowed operators 
in the $m-$basis is the following
\ba
&&\left<\Phi_{j_{1}; -j_{1}, -j_{1}}(z_{1})
\Phi_{j_{2}; m_{2}, {\bar m}_{2}}(z_{2})
\Phi_{j_{3}; m_{3}, {\bar m}_{3}}(z_{3})
\Phi_{j_{4}; m_{4}, {\bar m}_{4}}(z_{4})\right>\nonumber\\
&& \qquad\qquad \sim \;
V_{conf}\;\delta^{2}(m_{1}+m_{2}+m_{3}+m_{4})
\nonumber\\ &&
\quad\qquad\qquad
\times\; |z_{43}|^{2(\Delta_{1}+\Delta_{2}-\Delta_{3}-\Delta_{4})}
|z_{42}|^{-4\Delta_{2}}|z_{41}|^{2(\Delta_{2}+
\Delta_{3}-\Delta_{1}-\Delta_{4})}
|z_{31}|^{2(\Delta_{4}-\Delta_{1}-\Delta_{2}-\Delta_{3})}
\nonumber\\ &&
\quad\qquad\qquad\times\;\int dj\; C(j_{1},j_{2},j)\; B(j)^{-1}
C\left(j,j_{3},j_{4}\right)\nonumber\\ && 
\qquad\qquad\qquad\qquad\quad\times\;
[\;\Omega (j,j_{i},m_{i},{\bar m}_{i})\; +\; \lambda
\Omega (1-j,j_{i},m_{i},{\bar m}_{i})\; ]
\;
|z|^{2(\Delta_{j}-\Delta_{1}-\Delta_{2})}
\nonumber\\
&&\quad\qquad\qquad\qquad\quad\qquad\qquad\qquad\quad\qquad\qquad\qquad
\quad\qquad\qquad\qquad\quad
+\;\cdots \; ,
\label{corr}
\ea
where the ellipses denote higher orders in $z$.

We would now like to
discuss the more general case in which the highest weight operator 
$\Phi_{j_{1};-j_1,-j_1}$  is replaced by a global 
$SL(2,R)$ descendant, namely, we will consider the extension to
$m_{1}=-j_{1} -n_{1},~ {\bar m}_{1}=-j_{1} -{\bar n}_{1}$ with
$n_{1},{\bar n}_{1}=0,1,\cdots$,
which corresponds to an operator in the highest weight principal 
discrete representation ${\cal D}^{-}_{j_{1}}$ (see \cite{malda1} 
for notations  and conventions).

The procedure is analogous to that followed in 
\cite{becker} in the case 
of the three point function, and it involves acting on the correlator 
(\ref{corr}) with the lowering operator $J^+$ and making use of the
Baker-Campbell-Hausdorff formula (see \cite{becker} for details).
In the 
case of the three point function, the result can be expressed in terms of 
a sum over one running index.\footnote{See \cite{becker} for 
$m_{i}={\bar m}_{i}$ and \cite{satoh} for the more general case of the 
three point function with 
$m_{i}-{\bar m}_{i}\in {\bf Z}$.} In our case, however, there are
sums over two
holomorphic and two antiholomorphic indices, since
we are dealing with a four point function and
 $m_{i}-{\bar m}_{i}\in {\bf Z}$. In 
fact, performing these operations on (\ref{corr}), the following
generalization is obtained
\ba
&&\left<\Phi_{j_{1}; -j_{1}-n_{1}, -j_{1}-{\bar n}_{1}}(z_{1})
\Phi_{j_{2}; m_{2}, {\bar m}_{2}}(z_{2})
\Phi_{j_{3}; m_{3}, {\bar m}_{3}}(z_{3})
\Phi_{j_{4}; m_{4}, {\bar m}_{4}}(z_{4})\right>
\nonumber\\
&& =\; (-1)^{n_{1}+{\bar n}_{1}}\;\frac{\Gamma 
(2j_{1})^{2}}{\Gamma(j_{1}-m_{1})\Gamma(j_{1}-{\bar m}_{1})}
\nonumber\\ && \times \;
\sum_{n_{2},n_{3}=0}^{n_{1}}
\sum_{{\bar n}_{2},{\bar n}_{3}=0}^{{\bar n}_{1}}
{\cal G}_{n_{2},n_{3}}(j_{i},m_{i})\;
{\cal G}_{{\bar n}_{2},{\bar n}_{3}}(j_{i},{\bar m}_{i})
\nonumber\\
&&\qquad\times\;
\left<\Phi_{j_{1}; -j_{1}, -j_{1}}(z_{1})
\Phi_{j_{2}; m_{2}-n_{2}, {\bar m}_{2}-{\bar n}_{2}}(z_{2})
\Phi_{j_{3}; m_{3}-n_{3}, {\bar m}_{3}-{\bar n}_{3}}(z_{3})
\Phi_{j_{4}; m_{4}-n_{4}, {\bar m}_{4}-{\bar 
n}_{4}}(z_{4})\right>\; ,\nonumber\\
\label{corrr}
\ea
where $n_i \in$ {\bf Z}, $n_{1}=n_{2}+n_{3}+n_{4}$, ${\bar n}_{1}={\bar 
n}_{2}+{\bar n}_{3}+{\bar n}_{4}$, and we have defined 
\ba
&&{\cal G}_{n_{2},n_{3}}(j_{i},m_{i})\equiv\frac{1}{\Gamma(n_{2}+1)
\Gamma(n_{3}+1)}\; \frac{\Gamma(-j_{1}-m_{1}+1)}{\Gamma
(-j_{1}-m_{1}-n_{2}-n_{3}+1)}
\nonumber\\
&&\qquad\qquad\quad\times\;
\frac{\Gamma
(j_{2}-m_{2}+n_{2})\Gamma
(j_{3}-m_{3}+n_{3})\Gamma(j_{4}-j_{1}-m_{4}-m_{1}-n_{2}-n_{3})}
{\Gamma(j_{2}-m_{2})\Gamma(j_{3}-m_{3})\Gamma(j_{4}-m_{4})}
\; .\nonumber
\ea
The correlator in the r.h.s. of (\ref{corrr}) can be 
obtained by performing the replacements $m_{i}\rightarrow 
m_{i}-n_{i}$, ${\bar m}_{i}\rightarrow {\bar m}_{i}-{\bar n}_{i}$ 
($i=2,3,4$) in (\ref{corr}). In particular, note that this transforms 
\ba
&&\delta^{2}(m_{1}+m_{2}+m_{3}+m_{4})\nonumber\\ 
&&\qquad\qquad\longrightarrow\; 
\delta^{2}(m_{1}+(m_{2}-n_{2})+(m_{3}-n_{3})+(m_{4}+n_{2}+n_{3}-n_{1}))
\nonumber\\
&&\qquad\qquad\qquad =\; 
\delta^{2}((m_{1}-n_{1})+m_{2}+m_{3}+m_{4})\; ,
\nonumber
\ea
as expected for the correlator in the l.h.s. of (\ref{corrr}) since $m_{1}$ 
is lowered to $m_{1}-n_{1}$. 

Now plugging (\ref{corr}) into (\ref{corrr}) we get
\ba
&&\left<\Phi_{j_{1}; -j_{1}-n_{1}, -j_{1}-{\bar n}_{1}}(z_{1})
\Phi_{j_{2}; m_{2}, {\bar m}_{2}}(z_{2})
\Phi_{j_{3}; m_{3}, {\bar m}_{3}}(z_{3})
\Phi_{j_{4}; m_{4}, {\bar m}_{4}}(z_{4})\right>
\nonumber\\
&& \sim\; 
V_{conf}\;\delta^{2}((m_{1}-n_{1})+m_{2}+m_{3}+m_{4})\;\frac{\Gamma
(2j_{1})^{2}}{\Gamma(j_{1}-m_{1})\Gamma(j_{1}-{\bar m}_{1})}
\nonumber\\ &&
\quad
\times\; |z_{43}|^{2(\Delta_{1}+\Delta_{2}-\Delta_{3}-\Delta_{4})}
|z_{42}|^{-4\Delta_{2}}|z_{41}|^{2(\Delta_{2}+
\Delta_{3}-\Delta_{1}-\Delta_{4})}
|z_{31}|^{2(\Delta_{4}-\Delta_{1}-\Delta_{2}-\Delta_{3})}
\nonumber\\ &&
\quad\times\;
\sum_{n_{2},n_{3}=0}^{n_{1}}
\sum_{{\bar n}_{2},{\bar n}_{3}=0}^{{\bar n}_{1}}
{\cal G}_{n_{2},n_{3}}(j_{i},m_{i})\;
{\cal G}_{{\bar n}_{2},{\bar n}_{3}}(j_{i},{\bar m}_{i})\;
\int dj\; C(j_{1},j_{2},j)\; B(j)^{-1}
C\left(j,j_{3},j_{4}\right)
\nonumber\\
&&\qquad\quad\times\;
[\;\Omega (j,j_{i},m_{2}-n_{2},m_{3}-n_{3},
{\bar m}_{2}-{\bar n}_{2},{\bar m}_{3}-{\bar n}_{3})
\nonumber\\ &&\qquad\qquad
+\; \lambda
\Omega (1-j,j_{i},m_{2}-n_{2},m_{3}-n_{3},
{\bar m}_{2}-{\bar n}_{2},{\bar m}_{3}-{\bar n}_{3})\; ]
\; |z|^{2(\Delta_{j}-\Delta_{1}-\Delta_{2})}
\nonumber\\
&&\quad\qquad\qquad\qquad\quad\qquad\qquad\qquad\quad\qquad\qquad\qquad
\qquad\quad\quad
+\;\cdots \; ,
\label{desc}
\ea
where the ellipses denote higher orders in $z$.

A comment is in order. In the case of the three point function, the 
sum over one running index considered in \cite{becker} was extended to 
$\infty$ and then identified with the generalized hypergeometric function 
${_{3}F_{2}}$. However, in the present case we have not been able to 
reduce the sums in (\ref{desc}) to any elementary function, due to the 
much involved nature of the coefficient (\ref{oomega}) (it is even 
possible that such reduction is not at all practicable). 

Since the results (\ref{corr}) and (\ref{desc}) are written in the 
$m-$basis, we may now proceed to perform the spectral flow 
operation following the 
prescription (\ref{flow}) in order to obtain the winding conserving four 
point functions for states in arbitrary spectral flow sectors.
 For instance, applying the spectral flow operation to 
(\ref{corr}) gives
\ba
&&\left<\Phi^{w_{1},j_{1}=-m_{1}=-{\bar m}_{1}}_{J_{1},M_{1};{\bar J}_{1}
,{\bar M}_{1}}(z_{1})
\Phi^{w_{2},j_{2}}_{J_{2},M_{2};{\bar J}_{2}
,{\bar M}_{2}}(z_{2})
\Phi^{w_{3},j_{3}}_{J_{3},M_{3};{\bar J}_{3}
,{\bar M}_{3}}(z_{3})
\Phi^{w_{4},j_{4}}_{J_{4},M_{4};{\bar J}_{4}
,{\bar M}_{4}}(z_{4})\right>\nonumber\\
&& \qquad\qquad\sim \;
V_{conf}\;\delta^{2}(m_{1}+m_{2}+m_{3}+m_{4})~
z_{43}^{\Delta^{w_{1}}_{1}+\Delta^{w_{2}}_{2}-\Delta^{w_{3}}_{3}-
\Delta^{w_{4}}_{4}}
\nonumber\\ &&
\quad\qquad\qquad\times\; 
{\bar 
z}_{43}^{{\bar 
\Delta}^{w_{1}}_{1}+{\bar \Delta}^{w_{2}}_{2}-{\bar \Delta}^{w_{3}}_{3}-
{\bar \Delta}^{w_{4}}_{4}}
z_{42}^{-2\Delta^{w_{2}}_{2}}{\bar z}_{42}^{-2{\bar \Delta}^{w_{2}}_{2}}
z_{41}^{\Delta^{w_{2}}_{2}+
\Delta^{w_{3}}_{3}-\Delta^{w_{1}}_{1}-\Delta^{w_{4}}_{4}}
\nonumber\\ && \quad\qquad\qquad\times\;
{\bar z}_{41}^{{\bar \Delta}^{w_{2}}_{2}+
{\bar \Delta}^{w_{3}}_{3}-{\bar 
\Delta}^{w_{1}}_{1}-{\bar \Delta}^{w_{4}}_{4}}
z_{31}^{\Delta^{w_{4}}_{4}-\Delta^{w_{1}}_{1}-\Delta^{w_{2}}_{2}-
\Delta^{w_{3}}_{3}}
{\bar 
z}_{31}^{{\bar 
\Delta}^{w_{4}}_{4}-{\bar \Delta}^{w_{1}}_{1}-{\bar \Delta}^{w_{2}}_{2}-
{\bar \Delta}^{w_{3}}_{3}}
\nonumber\\ &&
\quad\qquad\qquad\times\;\int dj\; C(j_{1},j_{2},j)\; B(j)^{-1}
C\left(j,j_{3},j_{4}\right)~
z^{\Delta^{w}_{j}-\Delta^{w_{1}}_{1}-\Delta^{w_{2}}_{2}}
~{\bar z}^{{\bar 
\Delta}^{w}_{j}-{\bar \Delta}^{w_{1}}_{1}-{\bar \Delta}^{w_{2}}_{2}}
\nonumber\\ && \qquad\quad\qquad\qquad\qquad\qquad\times\;
[\;\Omega (j,j_{i},m_{i},{\bar m}_{i})\; +\; \lambda
\Omega (1-j,j_{i},m_{i},{\bar m}_{i})\; ]\;
\nonumber\\
&&\quad\qquad\qquad\qquad\quad\qquad\qquad\qquad\quad\qquad\qquad\qquad
\quad\qquad\qquad
+\;\cdots \; ,
\label{mbasis}
\ea
where $J_{i}=|M_{i}|=|m_{i}+\frac{k}{2}w_{i}|$, 
${\bar J}_{i}=|{\bar M}_{i}|=|{\bar m}_{i}+\frac{k}{2}w_{i}|$ are the 
spacetime 
conformal weights of the spectral flowed operators and they should be 
distinguished from the $j_{i}$ of the original 
unflowed states. 
Recall that this is a
winding conserving correlator with states in arbitrary spectral flow sectors 
up to the requirement 
$\sum_{i=1}^{4} w_{i}=0$. 

Similarly, applying the spectral flow operation to 
(\ref{desc}) one gets
\ba
&&\left<\Phi^{w_{1},j_{1}=-m_{1}-n_{1}=-{\bar
m}_{1}-{\bar n}_{1}}_{J_{1},M_{1};{\bar J}_{1}
,{\bar M}_{1}}(z_{1})
\Phi^{w_{2},j_{2}}_{J_{2},M_{2};{\bar J}_{2}
,{\bar M}_{2}}(z_{2})
\Phi^{w_{3},j_{3}}_{J_{3},M_{3};{\bar J}_{3}
,{\bar M}_{3}}(z_{3})
\Phi^{w_{4},j_{4}}_{J_{4},M_{4};{\bar J}_{4}
,{\bar M}_{4}}(z_{4})\right>
\nonumber\\
&& \sim\; 
V_{conf}\;\delta^{2}((m_{1}-n_{1})+m_{2}+m_{3}+m_{4})\;\frac{\Gamma
(2j_{1})^{2}}{\Gamma(j_{1}-m_{1})\Gamma(j_{1}-{\bar m}_{1})}
\nonumber\\ &&
\quad\times\;
z_{43}^{\Delta^{w_{1}}_{1}(n_{1})+\Delta^{w_{2}}_{2}-\Delta^{w_{3}}_{3}-
\Delta^{w_{4}}_{4}}
{\bar
z}_{43}^{{\bar
\Delta}^{w_{1}}_{1}({\bar n}_{1})+{\bar \Delta}^{w_{2}}_{2}-{\bar 
\Delta}^{w_{3}}_{3}-
{\bar \Delta}^{w_{4}}_{4}}
z_{42}^{-2\Delta^{w_{2}}_{2}}{\bar z}_{42}^{-2{\bar \Delta}^{w_{2}}_{2}}
\nonumber\\ && \quad\times\;
z_{41}^{\Delta^{w_{2}}_{2}+
\Delta^{w_{3}}_{3}-\Delta^{w_{1}}_{1}(n_{1})-\Delta^{w_{4}}_{4}}
{\bar z}_{41}^{{\bar \Delta}^{w_{2}}_{2}+
{\bar \Delta}^{w_{3}}_{3}-{\bar
\Delta}^{w_{1}}_{1}({\bar n}_{1})-{\bar \Delta}^{w_{4}}_{4}}
\nonumber\\ && \quad\times\;
z_{31}^{\Delta^{w_{4}}_{4}-\Delta^{w_{1}}_{1}(n_{1})-\Delta^{w_{2}}_{2}-
\Delta^{w_{3}}_{3}}
{\bar
z}_{31}^{{\bar
\Delta}^{w_{4}}_{4}-{\bar \Delta}^{w_{1}}_{1}({\bar n}_{1})-{\bar 
\Delta}^{w_{2}}_{2}-{\bar \Delta}^{w_{3}}_{3}}
\nonumber\\ &&
\quad\times\;
\sum_{n_{2},n_{3}=0}^{n_{1}}
\sum_{{\bar n}_{2},{\bar n}_{3}=0}^{{\bar n}_{1}}
{\cal G}_{n_{2},n_{3}}(j_{i},m_{i})\;
{\cal G}_{{\bar n}_{2},{\bar n}_{3}}(j_{i},{\bar m}_{i})\;
\int dj\; C(j_{1},j_{2},j)\; B(j)^{-1}
C\left(j,j_{3},j_{4}\right)
\nonumber\\
&&\qquad\qquad\qquad\quad\times\;
[\;\Omega (j,j_{i},m_{2}-n_{2},m_{3}-n_{3},
{\bar m}_{2}-{\bar n}_{2},{\bar m}_{3}-{\bar n}_{3})
\nonumber\\ &&\qquad\qquad\qquad\qquad\qquad
+\; \lambda
\Omega (1-j,j_{i},m_{2}-n_{2},m_{3}-n_{3},
{\bar m}_{2}-{\bar n}_{2},{\bar m}_{3}-{\bar n}_{3})\; ]
\nonumber\\ &&\qquad\qquad\qquad\qquad\qquad\qquad\times\;
z^{\Delta^{w}_{j}-\Delta^{w_{1}}_{1}(n_{1})-\Delta^{w_{2}}_{2}}
{\bar z}^{{\bar
\Delta}^{w}_{j}-{\bar \Delta}^{w_{1}}_{1}({\bar n}_{1})-{\bar 
\Delta}^{w_{2}}_{2}}\;
+\;\cdots \; ,
\label{corrrr}
\ea
where $\Delta_1^{w_1}(n_1)=\Delta_1^{w_1}+w_1n_1$ 
and $\bar\Delta_1^{w_1}(\bar n_1)=\bar \Delta_1^{w_1}+w_1\bar n_1$.
 
Having completed the analysis in the $m-$basis, 
our aim is to transform the four point functions
(\ref{mbasis}) and (\ref{corrrr}) back to 
the $x-$basis.
We follow a procedure analogous to that considered in \cite{malda3} 
in the case of two and three point functions.
It was shown that it is not necessary to compute the most general expression
since the $x-$basis correlators are the pole residue of the $m-$basis
results at 
$J_{i}=M_{i}$, ${\bar J}_{i}={\bar M}_{i}$, for a given spectral flowed 
state.\footnote{Actually, this 
procedure was applied in \cite{malda3} to the two point function
and the winding violating three point function. 
However, it may be shown
that it also gives the correct result 
for winding conserving three point functions comparing expressions 
in \cite{hmn} and \cite{satoh} (see the Appendix B).}
Similarly as in the case of the two point functions,
the pole here is in the divergent factor $V_{conf}$,
and (\ref{mbasis}) can then be 
interpreted as resulting from an $x-$basis expression of the form

\ba
&&\left<\Phi^{|w_{1}|,j_{1}}_{J_{1},{\bar J}_{1}}(x_{1},z_{1})
\Phi^{|w_{2}|,j_{2}}_{J_{2},{\bar J}_{2}}(x_{2},z_{2})
\Phi^{|w_{3}|,j_{3}}_{J_{3},{\bar J}_{3}}(x_{3},z_{3})
\Phi^{|w_{4}|,j_{4}}_{J_{4},{\bar J}_{4}}(x_{4},z_{4})
\right>\nonumber\\
&& \sim \;
x_{43}^{J_{1}+J_{2}-J_{3}-J_{4}}
{\bar x}_{43}^{{\bar J}_{1}+{\bar J}_{2}-{\bar J}_{3}-{\bar J}_{4}}
x_{42}^{-2J_{2}}{\bar x}_{42}^{-2{\bar J}_{2}}
\nonumber\\ && \times\;
x_{41}^{J_{2}+J_{3}-J_{1}-J_{4}}
{\bar x}_{41}^{{\bar J}_{2}+{\bar J}_{3}-{\bar J}_{1}-{\bar J}_{4}}
x_{31}^{J_{4}-J_{1}-J_{2}-J_{3}}
{\bar x}_{31}^{{\bar J}_{4}-{\bar J}_{1}-{\bar J}_{2}-{\bar J}_{3}}
\nonumber\\ && \times\;
z_{43}^{\Delta^{|w_{1}|}_{1}+\Delta^{|w_{2}|}_{2}-\Delta^{|w_{3}|}_{3}-
\Delta^{|w_{4}|}_{4}}
{\bar
z}_{43}^{{\bar
\Delta}^{|w_{1}|}_{1}+{\bar \Delta}^{|w_{2}|}_{2}-{\bar 
\Delta}^{|w_{3}|}_{3}-
{\bar \Delta}^{|w_{4}|}_{4}}
z_{42}^{-2\Delta^{|w_{2}|}_{2}}{\bar z}_{42}^{-2{\bar 
\Delta}^{|w_{2}|}_{2}}
\nonumber\\ && \times\;
z_{41}^{\Delta^{|w_{2}|}_{2}+
\Delta^{|w_{3}|}_{3}-\Delta^{|w_{1}|}_{1}-\Delta^{|w_{4}|}_{4}}
{\bar z}_{41}^{{\bar \Delta}^{|w_{2}|}_{2}+
{\bar \Delta}^{|w_{3}|}_{3}-{\bar
\Delta}^{|w_{1}|}_{1}-{\bar \Delta}^{|w_{4}|}_{4}}
\nonumber\\ && \times\;
z_{31}^{\Delta^{|w_{4}|}_{4}-\Delta^{|w_{1}|}_{1}-\Delta^{|w_{2}|}_{2}-
\Delta^{|w_{3}|}_{3}}
{\bar
z}_{31}^{{\bar
\Delta}^{|w_{4}|}_{4}-{\bar \Delta}^{|w_{1}|}_{1}-{\bar 
\Delta}^{|w_{2}|}_{2}-
{\bar \Delta}^{|w_{3}|}_{3}}
\nonumber\\ &&
\times\;\int dj\; C(j_{1},j_{2},j)\; B(j)^{-1}
C\left(j,j_{3},j_{4}\right)\;
[\;\Omega (j,j_{i},m_{i},{\bar m}_{i})\; +\; \lambda
\Omega (1-j,j_{i},m_{i},{\bar m}_{i})\; ]
\nonumber\\ && \qquad\qquad\qquad\qquad
\qquad\times\;
z^{\Delta^{|w|}_{j}-\Delta^{|w_{1}|}_{1}-\Delta^{|w_{2}|}_{2}}
{\bar z}^{{\bar
\Delta}^{|w|}_{j}-{\bar \Delta}^{|w_{1}|}_{1}-{\bar \Delta}^{|w_{2}|}_{2}}
x^{j-J_{1}-J_{2}}{\bar x}^{j-{\bar J}_{1}-{\bar J}_{2}}\nonumber\\ 
&&\qquad\qquad\quad\times\;
{\Bigg
\{}\;{\bigg
|}\;{_{2}F_{1}}(j-J_{1}+J_{2},j+J_{3}-J_{4},2j;x)\;{\bigg
|}^{2}\nonumber\\
&& \qquad
\qquad\qquad\quad
+\; {\hat \lambda} \;
{\bigg
|}\;x^{1-2j}{_{2}F_{1}}(1-j-J_{1}+J_{2},1-j+J_{3}-J_{4},2-2j;x)\;{\bigg
|}^{2}\;
{\Bigg \}}
\nonumber\\
&&\quad\qquad\qquad\qquad\quad\qquad\qquad\qquad\quad\qquad\qquad\qquad
\qquad\qquad\qquad
+\;\cdots \; ,
\label{xbasis}
\ea 
where the ellipses denote higher order terms in $z$. We have 
replaced $w_{i}\rightarrow |w_{i}|$ because in the $x-$basis the 
operators are labeled with positive winding number \cite{malda3} and
\ba
{\hat \lambda}=\;-\;\frac{\gamma(2j)^{2}\gamma(-J_{1}+J_{2}-j+1)
\gamma(J_{3}-J_{4}-j+1)}{(1-2j)^{2}\gamma(-J_{1}+J_{2}+j)
\gamma(J_{3}-J_{4}+j)}\; .
\label{llambda}
\ea

In order to determine the $x$ dependence of (\ref{xbasis})
we have used that  the Ward 
identities satisfied by
correlators involving either 
unflowed or spectral flowed fields in the $x-$basis 
are the same up to the replacements 
$\Delta_{i}\rightarrow \Delta_{i}^{w},\; j_{i}\rightarrow J_{i}$
for the spectral flowed fields \cite{hmn}. 
In addition, although 
the modified KZ and null vector equations to be satisfied by 
correlators involving spectral flowed fields in the $x-$basis were shown 
in \cite{hmn} to be iterative in the spins $J_{i}$, and their 
forms differ from the usual KZ 
and null vector equations for the unflowed case, 
at the lowest order in $z$ that we are 
considering here  the modified KZ equation actually reduces to that of the
unflowed case with the replacements $j_{i}\rightarrow J_{i}$, and
the iterative terms do not contribute. 
Therefore we expect that the
four point functions have the same 
dependence on the coordinates and  anharmonic ratios as 
(\ref{qa},\ref{qaa}), with the 
replacements mentioned above.

Similarly, the same procedure can be followed in order to transform 
(\ref{corrrr}) back to the $x-$basis and we get

\ba
&&\left<\Phi^{|w_{1}|,j_{1}}_{J_{1}(n_{1}),{\bar J}_{1}({\bar 
n}_{1})}(x_{1},z_{1})
\Phi^{|w_{2}|,j_{2}}_{J_{2},{\bar J}_{2}}(x_{2},z_{2})
\Phi^{|w_{3}|,j_{3}}_{J_{3},{\bar J}_{3}}(x_{3},z_{3})
\Phi^{|w_{4}|,j_{4}}_{J_{4},{\bar J}_{4}}(x_{4},z_{4})
\right>\nonumber\\
&& \sim \; \frac{\Gamma
(2j_{1})^{2}}{\Gamma(j_{1}-m_{1})\Gamma(j_{1}-{\bar m}_{1})}\;
x_{43}^{J_{1}(n_{1})+J_{2}-J_{3}-J_{4}}
{\bar x}_{43}^{{\bar J}_{1}({\bar n}_{1})+{\bar J}_{2}-{\bar J}_{3}-{\bar 
J}_{4}}
x_{42}^{-2J_{2}}{\bar x}_{42}^{-2{\bar J}_{2}}
\nonumber\\ && \times\;
x_{41}^{J_{2}+J_{3}-J_{1}(n_{1})-J_{4}}
{\bar x}_{41}^{{\bar J}_{2}+{\bar J}_{3}-{\bar J}_{1}({\bar n}_{1})-{\bar 
J}_{4}}
x_{31}^{J_{4}-J_{1}(n_{1})-J_{2}-J_{3}}
{\bar x}_{31}^{{\bar J}_{4}-{\bar J}_{1}({\bar n}_{1})-{\bar J}_{2}-{\bar 
J}_{3}}
\nonumber\\ && \times\;
z_{43}^{\Delta^{|w_{1}|}_{1}(n_{1})+\Delta^{|w_{2}|}_{2}-\Delta^{|w_{3}|}_{3}-
\Delta^{|w_{4}|}_{4}}
{\bar
z}_{43}^{{\bar
\Delta}^{|w_{1}|}_{1}({\bar n}_{1})+{\bar \Delta}^{|w_{2}|}_{2}-{\bar
\Delta}^{|w_{3}|}_{3}-
{\bar \Delta}^{|w_{4}|}_{4}}
z_{42}^{-2\Delta^{|w_{2}|}_{2}}{\bar z}_{42}^{-2{\bar
\Delta}^{|w_{2}|}_{2}}
\nonumber\\ && \times\;
z_{41}^{\Delta^{|w_{2}|}_{2}+
\Delta^{|w_{3}|}_{3}-\Delta^{|w_{1}|}_{1}(n_{1})-\Delta^{|w_{4}|}_{4}}
{\bar z}_{41}^{{\bar \Delta}^{|w_{2}|}_{2}+
{\bar \Delta}^{|w_{3}|}_{3}-{\bar
\Delta}^{|w_{1}|}_{1}({\bar n}_{1})-{\bar \Delta}^{|w_{4}|}_{4}}
\nonumber\\ && \times\;
z_{31}^{\Delta^{|w_{4}|}_{4}-\Delta^{|w_{1}|}_{1}(n_{1})-\Delta^{|w_{2}|}_{2}-
\Delta^{|w_{3}|}_{3}}
{\bar
z}_{31}^{{\bar
\Delta}^{|w_{4}|}_{4}-{\bar \Delta}^{|w_{1}|}_{1}({\bar n}_{1})-{\bar
\Delta}^{|w_{2}|}_{2}-
{\bar \Delta}^{|w_{3}|}_{3}}
\nonumber\\ &&\quad\times\;
\sum_{n_{2},n_{3}=0}^{n_{1}}
\sum_{{\bar n}_{2},{\bar n}_{3}=0}^{{\bar n}_{1}}
{\cal G}_{n_{2},n_{3}}(j_{i},m_{i})\;
{\cal G}_{{\bar n}_{2},{\bar n}_{3}}(j_{i},{\bar m}_{i})\;
\int dj\; C(j_{1},j_{2},j)\; B(j)^{-1}
C\left(j,j_{3},j_{4}\right)\nonumber\\
&&\qquad\qquad\qquad\quad\times\;
[\;\Omega (j,j_{i},m_{2}-n_{2},m_{3}-n_{3},
{\bar m}_{2}-{\bar n}_{2},{\bar m}_{3}-{\bar n}_{3})
\nonumber\\ &&\qquad\qquad\qquad\qquad\qquad
+\; \lambda
\Omega (1-j,j_{i},m_{2}-n_{2},m_{3}-n_{3},
{\bar m}_{2}-{\bar n}_{2},{\bar m}_{3}-{\bar n}_{3})\; ]
\nonumber\\ && \qquad\qquad
\qquad\times\;
z^{\Delta^{|w|}_{j}-\Delta^{|w_{1}|}_{1}(n_{1})-\Delta^{|w_{2}|}_{2}}
{\bar z}^{{\bar
\Delta}^{|w|}_{j}-{\bar \Delta}^{|w_{1}|}_{1}({\bar n}_{1})-{\bar 
\Delta}^{|w_{2}|}_{2}}
x^{j-J_{1}(n_{1})-J_{2}}{\bar x}^{j-{\bar J}_{1}({\bar n}_{1})-{\bar 
J}_{2}}\nonumber\\
&&\qquad\qquad\quad\times\;
{\Bigg
\{}\;{\bigg
|}\;{_{2}F_{1}}(j-J_{1}(n_{1})+J_{2},j+J_{3}-J_{4},2j;x)\;{\bigg
|}^{2}\nonumber\\
&& \qquad
\qquad\quad
+\; {\hat \lambda}(n_{1}) \;
{\bigg
|}\;x^{1-2j}{_{2}F_{1}}(1-j-J_{1}(n_{1})
+J_{2},1-j+J_{3}-J_{4},2-2j;x)\;{\bigg
|}^{2}\;
{\Bigg \}}
\nonumber\\
&&\quad\qquad\qquad\qquad\quad\qquad\qquad\qquad\quad\qquad\qquad\qquad
\quad\qquad\qquad
+\;\cdots \; ,
\label{xxbasis}
\ea
where $J_{1}(n_{1})=|-j_1-n_1+\frac k2 w_1|$.

Therefore we have extended the result (\ref{qa},\ref{qaa}) for the 
four point function of unflowed operators in the $SL(2,R)$ WZW model, to all 
winding conserving four 
point functions for states in arbitrary spectral flow sectors. We have 
obtained results both in the $m-$ and the $x-$basis, 
with the simplifying assumption that at least one operator is in 
the spectral flow image of ${\cal D}_{j}^{-}$.

Notice that, while the $m-$basis expressions (\ref{mbasis}) and (\ref{corrrr}) 
remain valid
 for all winding conserving four point functions, including in particular
the case in which all the external operators are unflowed, the results
(\ref{xbasis}) and (\ref{xxbasis}) in the
$x-$basis do not hold when all the external states are unflowed.
This is consistent with the fact that in the $m-$basis, all $N-$point 
functions are the same, for a given $N$,  up to some free boson correlator 
\cite{malda1}\cite{malda3}\cite{ribault} which only modifies the $z_i$ 
dependence. 
On the other hand, the procedure we have followed to transform 
from the $m-$ to the 
$x-$basis, by evaluating the pole residue at $J=M$,
${\bar J}={\bar M}$, requires at least one  
spectral flowed state in the correlator.\footnote{In fact there should be 
at least
two spectral flowed operators in the winding conserving case.}
Consequently (\ref{xbasis}) and (\ref{xxbasis}) result in this case, whereas 
(\ref{qa},\ref{qaa}) hold for four unflowed operators.

We would like to comment on the higher order contributions to the 
expansion in $z$. As in the case of 
the 
four point function of unflowed operators \cite{tesch2}, we expect them to 
be determined once the lowest order is given as the initial condition. 
This works in two different but equivalent ways. The first one is that 
higher orders in the spectral flowed case are fixed by first 
iteratively determining the higher orders in (\ref{expan}) starting from 
(\ref{qaa}), and then performing the spectral flow operation to the 
result, in a similar fashion as we have done
 here to the lowest order contribution.  
Alternatively one may determine the higher order contributions using 
modified KZ equations for amplitudes involving spectral flowed states, 
starting from (\ref{mbasis}$-$\ref{xxbasis}) as the initial conditions. In the 
$m-$ and $\mu -$basis, such modified KZ equations were discussed in 
\cite{ribault}.
In the $x-$basis, modified KZ 
equations were computed in \cite{hmn} and the determination of the higher 
order contributions was also discussed in the one-unit violating case.

Finally, one important application of our results
would be to investigate the structure of the factorization of 
(\ref{mbasis}$-$\ref{xxbasis})
in order to establish the consistency of 
string theory on $AdS_{3}$ 
and verify the winding violation pattern suggested in \cite{malda3}. 
We hope to tackle this problem
in the future. 

{\large{\bf Acknowledgments}}:
We would like to thank E. Herscovich for many helpful discussions and J. 
Maldacena for reading the manuscript. This work 
was supported by CONICET, Universidad de Buenos Aires and ANPCyT.

\vspace*{0.5cm}
 {\large {\bf Appendix A. Useful formulae}}
\vspace*{0.5cm}

The following identities for the hypergeometric functions can be found 
e.g. in \cite{luke}

\ba
\int_{0}^{1}du\; u^{\alpha}(1-u)^{\beta}{_{2}F_{1}}(a,b,c;\lambda u)=
\frac{\Gamma(\alpha +1)\Gamma(\beta +1)}{\Gamma(\alpha +\beta +2)}\;
{_{3}F_{2}}(a,b,\alpha +1;c,\alpha+\beta+2;\lambda)\; .\nonumber\\
\label{m1}
\ea
\ba
&&\int_{0}^{1}du\; 
u^{\alpha}(1-u)^{\beta}{_{3}F_{2}}(a,b,c;d,e;u)\nonumber\\ &&
\qquad\qquad\qquad =\;
\frac{\Gamma(\alpha +1)\Gamma(\beta +1)}{\Gamma(\alpha +\beta +2)}\;
{_{4}F_{3}}(a,b,c,\alpha +1;d,e,\alpha+\beta+2;1)\; .
\label{m2}
\ea
\ba
&&{_{3}F_{2}}(a,b,c;d,e;u)\nonumber\\
&&=\; \frac{\Gamma(d)\Gamma(e)}{\Gamma(a)\Gamma(b)\Gamma(c)}\;
{\Bigg [}\;
\frac{\Gamma(a)\Gamma(b-a)\Gamma(c-a)}{\Gamma(d-a)\Gamma(e-a)}\;
(-u)^{-a}\nonumber\\
&&\qquad\qquad\qquad\qquad\qquad\times\;
{_{3}F_{2}}\left(a-d+1,a-e+1,a;a-b+1,a-c+1;\frac{1}{u}\right)
\nonumber\\
&& \qquad\qquad\qquad\qquad + \;
\frac{\Gamma(b)\Gamma(a-b)\Gamma(c-b)}{\Gamma(d-b)\Gamma(e-b)}\;
(-u)^{-b}\nonumber\\
&&\qquad\qquad\qquad\qquad\qquad\times\;
{_{3}F_{2}}\left(b-d+1,b-e+1,b;b-a+1,b-c+1;\frac{1}{u}\right)
\nonumber\\
&& \qquad\qquad\qquad\qquad + \;
\frac{\Gamma(c)\Gamma(a-c)\Gamma(b-c)}{\Gamma(d-c)\Gamma(e-c)}\;
(-u)^{-c}\nonumber\\
&&\qquad\qquad\qquad\qquad\qquad\times\;
{_{3}F_{2}}\left(c-d+1,c-e+1,c;c-a+1,c-b+1;\frac{1}{u}\right)
\;{\Bigg ]}\; .\nonumber\\
\label{m3}
\ea

{\large {\bf Appendix B. Three point functions}}
\vspace*{0.5cm}
\\
In this appendix we transform  the three point function including two 
$w=1$ 
operators, computed in \cite{hmn}, from the $x-$basis to the $m-$basis. 
We
verify that the result equals the three point function for unflowed 
operators in the $m-$basis computed in \cite{satoh}. This may be 
considered a check not only on the expressions in \cite{hmn}\cite{satoh}, 
but also on the claim that the coefficient of all 
winding conserving correlators is the same in the $m-$basis
(for a given number of external 
states) \cite{malda1}\cite{malda3}\cite{ribault}. 

The 
three point function 
including two $w=1$ operators in the $x-$basis is   \cite{hmn} 
\ba
&&\left<\Phi^{w
=1,j_{1}}_{J_{1},{\bar J}_{1}}(x_{1},z_{1})\Phi_{J_{2},{\bar J}_{2}}^{w=1,
j_2}(x_{2},z_{2})
\Phi_{j_{3}}(x_{3},z_{3})\right>
\nonumber\\ && \quad
 \sim\;\frac{1}{V^{2}_{conf}}\; B(j_1)B(j_2)C\left (\frac k2-j_1,
\frac k2-j_2, j_3\right )
W(j_1,j_2,j_3,m_1,m_2,\bar m_1,\bar m_2)\nonumber\\
&&\quad \times ~
x_{12}^{j_3-J_1-J_2}
\bar x_{12}^{j_3-\bar J_1-\bar J_2}
x_{13}^{J_2-J_1-j_3}
\bar x_{13}^{\bar J_2-\bar J_1-j_3}
x_{23}^{J_1-J_2-j_3}
\bar x_{23}^{\bar J_1-\bar J_2-j_3}
\nonumber\\
&&\quad \times ~
z_{12}^{\Delta_3-\Delta_1^{w=1}-\Delta_2^{w=1}}
\bar z_{12}^{\Delta_3-\bar \Delta_1^{w=1}-\bar \Delta_2^{w=1}}
z_{13}^{\Delta_2^{w=1}-\Delta_1^{w=1}-\Delta_3}
\bar z_{13}^{\bar \Delta_2^{w=1}-\bar \Delta_1^{w=1}-\Delta_3}\nonumber\\
&&\quad \times ~
z_{23}^{\Delta_1^{w=1}
-\Delta_2^{w=1}-\Delta_3}
\bar z_{23}^{\bar \Delta_1^{w=1}-\bar \Delta_2^{w=1}-\Delta_3}\; ,
\label{3pww}
\ea
where
\ba
&& W(j_1,j_2,j_3,m_1,m_2,\bar m_1,\bar m_2) =
\int d^{2}u\; d^{2}v\; u^{j_{1}-m_{1}-1}{\bar u}^{j_{1}-{\bar m}_{1}-1}
v^{j_{2}-m_{2}-1}{\bar v}^{j_{2}-{\bar m}_{2}-1}\nonumber\\ &&
\qquad\qquad\qquad\qquad \qquad\qquad\quad
\times \;
|1-u|^{2(j_{2}-j_{1}-j_{3})}|1-v|^{2(j_{1}-j_{2}-j_{3})}
|u-v|^{2(j_{3}-j_{1}-j_{2})}\; ,
\nonumber
\ea
was computed in \cite{fh}, but the explicit result is 
not relevant for our purposes here.

In order to transform this expression to the $m-$basis, we 
use (\ref{max}) and evaluate the residue of the pole at $J_{2}=-M_{2}$, 
${\bar J}_{2}=-{\bar M}_{2}$ in the $x_{2}$ integral, $i.e.$ we
perform the following operation to (\ref{3pww})
\ba
\int d^2 x_{1}\; d^2 x_{3}\; x_{1}^{J_{1}-M_{1}-1}
{\bar x}_{1}^{{\bar J}_{1}-{\bar M}_{1}-1}
x_{3}^{j_{3}-m_{3}-1}
{\bar x}_{3}^{j_{3}-{\bar m}_{3}-1}
\lim_{x_{2},{\bar x}_{2}\rightarrow\infty} \; x_{2}^{2J_{2}}
{\bar x}_{2}^{2{\bar J}_{2}}\; .
\nonumber
\ea
Thus we find that the $x-$dependent part 
in (\ref{3pww}) 
contributes the following factor
\beq
V_{conf}\;\delta^{2}(M_{1}+M_{2}+m_{3})\;
\frac{\Gamma(j_{3}-m_{3})\Gamma(J_{2}-J_{1}-j_{3}+1)
\Gamma({\bar J}_{1}-{\bar J}_{2}+{\bar m}_{3})}
{\Gamma(-j_{3}+{\bar m}_{3}+1)\Gamma({\bar J}_{1}-{\bar
J}_{2}+j_{3})\Gamma(J_{2}-J_{1}-m_{3}+1)}\; ,
\label{990}
\eeq
up to some $k$-dependent ($j$ independent) coefficient.
Considering that $J_{1}=M_{1}
=m_1+k/2$, $J_{2}=-M_{2}=-m_2+k/2$, $M_{1}+M_{2}+m_{3}=0$ and
$m_3-\bar m_3 \in$ {\bf Z}, this can be reduced to
\beq
V^{2}_{conf}\; \delta^{2}(m_{1}+m_{2}+m_{3})\; .
\label{994}
\eeq
Therefore, recalling the identity \cite{hmn}
\beq
B(j_1)B(j_2)C\left (\frac k2-j_1, \frac k2-j_2, j_3\right )\sim
C(j_{1},j_{2},j_{3})\; ,
\label{995}
\eeq
we obtain the following
expression for the winding conserving ($w_1=-w_2=1, w_3=0$)
three point function 
in the $m$-basis
\ba
&&\left<\Phi^{w
=1,j_{1}}_{J_{1},M_{1};{\bar J}_{1},{\bar
M}_{1}}(z_{1})\Phi^{w
=-1,j_{2}}_{J_{2},M_{2};{\bar J}_{2},{\bar
M}_{2}}(z_{2})
\Phi_{j_{3};m_{3},{\bar m}_{3}}(z_{3})\right>
\nonumber\\ && \quad
 \sim
\; \delta^{2}(m_{1}+m_{2}+m_{3})\; C(j_{1},j_{2},j_{3})\;
W(j_1,j_2,j_3,m_1,m_2,\bar m_1,\bar m_2)\nonumber\\
&&\quad \times ~
z_{12}^{\Delta_3-\Delta_1^{w=1}-\Delta_2^{w=-1}}
\bar z_{12}^{\Delta_3-\bar \Delta_1^{w=1}-\bar \Delta_2^{w=-1}}
z_{13}^{\Delta_2^{w=-1}-\Delta_1^{w=1}-\Delta_3}
\bar z_{13}^{\bar \Delta_2^{w=-1}-\bar \Delta_1^{w=1}-\Delta_3}\nonumber\\
&&\quad \times ~
z_{23}^{\Delta_1^{w=1}
-\Delta_2^{w=-1}-\Delta_3}
\bar z_{23}^{\bar \Delta_1^{w=1}-\bar \Delta_2^{w=-1}-\Delta_3}\; .
\label{3pwwm}
\ea

This expression coincides with the three point function of unflowed
operators computed in \cite{satoh} up to the powers of $z_{ij}$, which 
have to be transformed according to (\ref{flow}).

\end{document}